\documentclass[aps,pra,reprint,nofootinbib,twocolumn,superscriptaddress,showpacs,showkeys,longbibliography]{revtex4-1}

\usepackage{eurosym}
\usepackage{amsmath,amssymb,amstext}
\usepackage{ulem}
\usepackage[usenames,dvipsnames]{color}
\usepackage{graphicx}
\usepackage{braket}
\usepackage{natbib}
\usepackage{comment}
\usepackage{dcolumn}
\usepackage[english]{babel}
\usepackage{wasysym}
\usepackage[colorlinks,bookmarks=false,citecolor=blue,linkcolor=red,urlcolor=blue]{hyperref}

\begin{document}

\title{Role of interaction in the binding of two Spin-orbit Coupled Fermions}

\author{Chong Ye}
\affiliation{Graduate School, China Academy of Engineering Physics, Beijing 100193,
China}
\affiliation{National Laboratory of Science and Technology on Computational Physics,
Institute of Applied Physics and Computational Mathematics, Beijing 100088,
China}

\author{Jie Liu}
\affiliation{National Laboratory of Science and Technology on Computational Physics,
Institute of Applied Physics and Computational Mathematics, Beijing 100088,
China}
\affiliation{HEDPS, CAPT, and CICIFSA MoE, Peking University, Beijing 100871,
China}

\author{Li-Bin Fu}\email{lbfu@gscaep.ac.cn}
\affiliation{Graduate School, China Academy of Engineering Physics, Beijing 100193,
China}
\begin{abstract}
 We investigate role of an attractive s-wave interaction with positive scattering length in the binding of two spin-orbit coupled fermions in the vacuum and on the top of a Fermi sea in the single impurity system, motivated by current interests in exploring
 exotic binding properties in the appearance of spin-orbit couplings. For weak spin-orbit couplings where the density of states is not significantly altered, we analytically show that the high-energy states become more important in determining the binding energy when the scattering length decreases. Consequently, tuning the interaction gives rise to a rich behavior, including a zigzag of the momentum of the bound state or inducing transitions among the meta-stable states. By exactly solving the two-body quantum mechanics for a spin-orbit coupled Fermi mixture of $^{40}$K-$^{40}$K-$^{6}$Li, we demonstrate that our analysis can also apply to the case when the density of states is significantly modified by the spin-orbit coupling. Our findings pave a way for understanding and controlling the binding of fermions in the presence of spin orbit couplings.
\end{abstract}

\pacs{03.65.Ge, 71.70.Ej, 67.85.Lm}
\maketitle
\section{Introduction}
In ultracold physics, many schemes have been proposed to generated various types of synthetic
spin-orbit couplings (SOC) by controlling atom-light interaction \cite{TSOC1}. In 2011, I. B. Spielman's group in NIST had generated an equal weight combination of Rashba-type and Dresselhaus-type SOC in $^{87}$Rb \cite{ESOCB}. Afterwards, SOC has triggered a great amount of experimental interest \cite{ESOCB1,ESOCF1,ESOCF}. In the appearance of SOC, the ultracold atomic gases have been altered
dramatically \cite{ZH1,ZH2,ISB}.

One basic issue is the binding of two spin-orbit coupled fermions
in the vacuum \cite{PRB.83.094515,PRL.107.195305,
PRB.84.014512,PRL.107.195304,PRA.87.033629,
PRA.88.033609,PRA.89.063610,PRL.112.195301} where SOC has given rise to
the change of binding energy and the appearance of finite-momentum
dimer bound states. Another relevant issue is the binding of two fermions on the top of a Fermi sea
(the molecular state)
for the case where a single impurity is immersed in a noninteracting
Fermi gas \cite{PRL.112.195301,c9,c10,c11,c12,c13}.
In the appearance of SOC, the center-of-mass (c.m.) momentum of the molecular state becomes finite
\cite{PRL.112.195301,c13}. All of these can be understood from the perspective of two-body
quantum mechanics. It general contains three components:
the threshold energy associated with the c.m. momentum, the density of states,
and the interacting strength. For extremely weak attractive interaction, changes of
two-body properties under SOC came from the different threshold behavior of the
density of states \cite{PRB.83.094515,PRB.84.014512,PRA.88.033609}. However,
in the strong interacting regime, the binding of two fermions presents
a rich behavior \cite{PRL.112.195301,PRA.87.033629} such as the variation of
the c.m. momentum and the competition between two meta-stable states with the tuning of
interacting strength. These
phenomena can not simply owe to the threshold behavior of the density of states.
Therefore, the mechanism as to how all these
three components cooperate with each other in determining the novel two-body
properties is pressing needed.
The establishment of such a comprehensive picture will shed light on ongoing
explorations of the intriguing behavior of spin-orbit coupled Fermi gases \cite{PRB.83.094515,PRL.107.195305,
PRB.84.014512,PRL.107.195304,PRA.87.033629,
PRA.88.033609,PRA.89.063610,PRL.112.195301}.
Below, we report a theoretical
contribution to address this issue, which also allows predictions of new phenomena.

We investigate the two-body quantum mechanisms of the binding of two spin-orbit coupled fermions in the vacuum and on the top of a Fermi sea in
the single impurity Fermi gas. We consider an attractive s-wave interaction with positive scattering length,
the strength of which can be tuned in a wide range via a Feshbach resonance \cite{RMP.82.1225}.
From Sec. II to Sec. IV, we give analyses which do not dependent on the concrete type of SOC. In Sec. II, by decomposing the
two-body energy (molecular energy) into the threshold
energy and the binding energy, both of which depend on the c.m. momentum of two fermions, we establish a direct relation
between the interaction, the density of states, and the binding energy. In Sec. III, with the first-order perturbation analysis in
the weak SOC limit, we reveals that the low-energy states play a decisive role in determining the binding energy when
the scattering length is large, in contrast to the small scattering length case where the high-energy states can dominate.
This allows us to elucidate the mechanism
underlying interesting phenomena such as a zigzag behavior of the two-body ground state momentum and the competition between
two meta-stable states in Sec. IV.  In Sec. V, we illustrate our analysis with an interacting Fermi mixture of $^{40}$K-$^{40}$K-$^{6}$Li with $^{40}$K
containing an ($\alpha k_x\sigma_z+h\sigma_x$)-type SOC, which can be realized by
the state of the art experimental techniques using cold atoms \cite{ESOCF,FM}. Remarkably, by exactly
solving the two-body problem for this system, we show that our analysis affords insights into the main properties of the binding of two spin-orbit coupled fermions, even when the density of states is significantly altered by SOC. Our findings reveal the role
of interaction in the binding of two spin-orbit coupled fermions and allow
deep physical understandings of the rich two-body properties in the presence of SOC.

\section{Binding of two Fermions with SOC}

We consider two different spin-orbit coupled fermionic species in three dimensions (3D) at zero temperature: atom A (B) has $N_a$ ($N_b$) components, with the corresponding non-interacting Hamiltonian $H_a$ ($H_b$). We consider an attractive s-wave contact interaction with positive scattering length between the two fermionic species as described by
\begin{align}
H_{int}=\frac{U}{V}\sum_{\mathbf{Q,k,k^{\prime}}}
a^{\dagger}_{\mathbf{k},l_{0}}b^{\dagger}_{\mathbf{Q-k},m_{0}} a_{\mathbf{k^{\prime}},l_{0}}b_{\mathbf{Q-k^{\prime}},m_{0}},
\end{align}
with $\mathbf{Q}$ the c.m. momentum of two scattering fermions. Here $a^\dag_{{\bf k},l_0}$ ($b^\dag_{{\bf k},m_0}$) denotes the creation operator of a SOC-free atom A (B) in the $l_0$-th ($m_0$-th) spin component with momentum ${\bf k}$, $U$ is the bare interaction, and $V$ is the quantization volume. The total Hamiltonian is thus $H=H_a+H_b+H_{\textrm{int}}$.

With this
Hamiltonian, we address to the binding of two fermionic atoms A and B (i) in the vacuum \cite{PRB.83.094515,PRL.107.195305,
PRB.84.014512,PRL.107.195304,PRA.87.033629,
PRA.88.033609,PRA.89.063610,PRL.112.195301} and (ii) on the top of a non-interacting Fermi sea of atoms $A$ in the situation where a single impurity of $B$ is immersed in a non-interacting Fermi gas of $A$ \cite{c9,c10,c11,c12,c13}. The ansatz wave function of the two-body bound state and the molecular state can be expressed in a general form
\begin{align}
|\Psi_{\mathbf{Q}}\rangle=\sum_{i,j}\sum^{\prime}_{\mathbf{k}}\psi^{i,j}_{\mathbf{Q,k}}
\alpha^{\dagger}_{\mathbf{k},i}\beta^{\dagger}_{\mathbf{Q-k},j}
|\varnothing\rangle,
\end{align}
where $\mathbf{Q}$ is the c.m. momentum of two particles. For (i), $|\varnothing\rangle$ is the vacuum state and
the summation $\sum^{\prime}_{\mathbf{k}}$ includes all the states. For (ii), $|\varnothing\rangle$ is the non-interacting spin-orbit coupled Fermi sea of $A$ and the summation $\sum^{\prime}_{\mathbf{k}}$ excludes the states below the Fermi surfaces, reflecting the effect of Pauli blocking. Here, $\alpha^{\dagger}_{\mathbf{k},i}=\sum_{i^{\prime}}
\lambda^{i,i^{\prime}}_{\mathbf{k}}a^{\dagger}_{\mathbf{k},i^{\prime}}$
($\beta^{\dagger}_{\mathbf{k},i}=\sum_{i^{\prime}}
\eta^{i,i^{\prime}}_{\mathbf{k}}b^{\dagger}_{\mathbf{k},i^{\prime}}$) is the creation operator of an atom A (B) in the $i$-th eigen-state of Hamiltonian $H_a$ ($H_b$) with momentum $\mathbf{k}$ and energy $\varepsilon^{a}_{\mathbf{k},i}$ ($\varepsilon^{b}_{\mathbf{k},i}$) and $\psi^{i,j}_{\mathbf{Q,k}}$ denotes the variational coefficient. The coefficients $\lambda^{i,i^{\prime}}_{\mathbf{k}}$ and $\eta^{i,i^{\prime}}_{\mathbf{k}}$ are fixed by SOC.
Solving the eigen-equation $H|\Psi_{\mathbf{Q}}\rangle
=E_{\mathbf{Q}}|\Psi_{\mathbf{Q}}\rangle$ gives
\begin{equation}
\psi^{i,j}_{\mathbf{Q,k}}=\frac{(\lambda^{i,l_0}_{\mathbf{k}}\eta^{j,m_{0}}_{\mathbf{Q-k}})^{\ast}}
{E_{\mathbf{Q}}-E^{ij}_{\mathbf{Q,k}}}\frac{U}{V}\sum^{\prime}_{\mathbf{k}^{\prime},i^{\prime},j^{\prime}}
\psi^{i^{\prime},j^{\prime}}_{\mathbf{Q,k^{\prime}}}\lambda^{i^{\prime},l_0}_{\mathbf{k^{\prime}}}\eta^{j^{\prime},m_{0}}_{\mathbf{Q-k^{\prime}}}, \label{Eq1}
\end{equation}
with $E^{ij}_{\mathbf{Q},\mathbf{k}}=\varepsilon^{a}_{\mathbf{k},i}+\varepsilon^{b}_{\mathbf{Q-k},j}$. Rearranging Eq. (\ref{Eq1}), we obtain a self-consistent equation for two-body energy (molecular energy) $E_{\mathbf{Q}}$ in the momentum-space representation, i.e.,
\begin{align}\label{Eq2}
\frac{1}{U}=\frac{1}{V}\sum_{i,j}{\sum_{\mathbf{k}}}^{\prime}
\frac{|\lambda^{i,l_{0}}_{\mathbf{k}}|^2 |\eta^{j,m_{0}}_{\mathbf{Q-k}}|^2}{E_{\mathbf{Q}}-E^{ij}_{\mathbf{Q},\mathbf{k}}}.
\end{align}

A key step of our treatment next constitutes a decomposition of  $E_{\mathbf{Q}}$: Defining the threshold energy associated with the
c.m. momentum $\mathbf{Q}$ by $E_{\textrm{th}}^{\mathbf{Q}}\equiv \textrm{min}_{i,j,\mathbf{k}}\{E^{ij}_{\mathbf{Q},\mathbf{k}}\}$, we write $E^{ij}_{\mathbf{Q}}=E^{\mathbf{Q}}_{th}+\varepsilon$.
The rest of the two-body energy (molecular energy) is therefore $E^{sc}_{\mathbf{Q}}\equiv E_{\mathbf{Q}}-E^{\mathbf{Q}}_{th}$. While $E_{\textrm{th}}^{\mathbf{Q}}$ is only affected by SOC,  $E^{sc}_{\mathbf{Q}}$ encodes the effect of interaction.  Such decomposition of $E_{\mathbf{Q}}$ in terms of $E^{sc}_{\mathbf{Q}}$ and $E_{\textrm{th}}^{\mathbf{Q}}$, as we shall see, allows a transparent correspondence to the SOC-free counterpart. Following a standard procedure, we obtain the self-consistent
equation for $E^{sc}_{\mathbf{Q}}$ in the energy domain of $\varepsilon$ as in Ref.\cite{RdNC.31.247}
\begin{align}\label{Eq3}
\int^{\infty}_{0}
\frac{\gamma^{\varepsilon}_{\mathbf{Q}}d\varepsilon}{E^{sc}_{\mathbf{Q}}-\varepsilon}=\frac{1}{U}.
\end{align}
Here, $\gamma^{\varepsilon}_{\mathbf{Q}}$ is defined by
\begin{align}\label{Eq4}
\gamma^{\varepsilon}_{\mathbf{Q}}=\sum_{i}\sum_{j}
\int^{\prime}
|\lambda^{i,l_{0}}_{\mathbf{k}}|^2 |\eta^{j,m_{0}}_{\mathbf{Q-k}}|^2|J|d\nu d\mu,
\end{align}
which describes the density of states in 3D\cite{foot}. For (i), the integration $\int^{\prime} d\nu d\mu$ includes all
the states. For (ii), the integration $\int^{\prime} d\nu d\mu$ excludes the states below the Fermi surfaces. In Eq. (\ref{Eq4}), $\mu$ and $\nu$ label the degrees of freedom other than $\varepsilon$, and $J$ denotes the standard Jacobian. These formulas can be also easily adapted to describing the binding of two homo-nuclear fermions where $A$ and $B$ are the same fermionic species.

Equation (\ref{Eq3}) establishes a direct relation between the interaction $U$, the density of states $\gamma^{\varepsilon}_{\mathbf{Q}}$, and $E^{sc}_{\mathbf{Q}}$. Intuition behind it can be gained in the limit of vanishing SOC in case (i), where $E^{\mathbf{Q}}_{th}=\mathbf{Q}^2/(2m_\mu)$ with $m_\mu$ the reduced mass of two fermions, and $\gamma^{\varepsilon}_{\mathbf{Q}}=\gamma^{\varepsilon}_{0}=2\sqrt{2m_{\mu}\varepsilon}$. Then, $E^{sc}_{\mathbf{Q}}$ is independent of $\mathbf{Q}$ as ensured by Eq. (\ref{Eq3}), and can be identified as $E^{sc}_{\mathbf{Q}}\equiv \varepsilon_b=-1/(2m_{\mu}a^2_s)$ [$\hbar\equiv 1$] with $a_s>0$ the s-wave scattering length, i.e., the binding energy at rest. In this case, Eq. (\ref{Eq3}) reduces to, in the momentum space representation, the well known renormalization equation
for two scattering particles, i.e.,
\begin{align}
\frac{1}{U}=\frac{m_{\mu}}{2\pi a_{s}}-\frac{1}{V}\sum_{\mathbf{k}}
\frac{2m_\mu}{\mathbf{k}^2}.
 \end{align}
Equation (\ref{Eq3}) thus extends the standard prescription for two interacting fermions to the presence of SOC, where $E_{\mathbf{Q}}^{sc}$ is the counterpart of the binding energy $\varepsilon_b$.

\section{role of interaction}
Based on above treatment, below we elucidate how the interaction cooperates with the effect of SOC in determining the behavior of $E^{sc}_{\mathbf{Q}}$, when the interaction strength $a^{-1}_s$ is tuned in a wide range via Feshbach resonance \cite{RMP.82.1225}. To compare to the SOC-free case, we introduce the quantity $\xi_{\mathbf{Q}}\equiv E^{sc}_{\mathbf{Q}}-\varepsilon_{b}$. For weak SOC that does not significantly alter the density of states, the leading term of $\xi_{\mathbf{Q}}$ can be derived from Eq. (\ref{Eq3}) as \cite{Sup1}:
\begin{align}\label{Eq5}
\xi_{\mathbf{Q}}=
-\Big[\int^{\infty}_{0}\frac{\gamma^{\varepsilon}_{\mathbf{Q}}}{(\varepsilon-\varepsilon_{b})^2}d\varepsilon\Big]^{-1}
\int^{\infty}_{0}\frac{\gamma^{\varepsilon}_{\mathbf{Q}}-\gamma^{\varepsilon}_{0}}{\varepsilon-\varepsilon_{b}}
d\varepsilon.
\end{align}
Here we have ignored the modification of
the renormalization relation by SOC \cite{PRA.85.022705,PRA.86.053608,PRA.86.042707,PRA.87.032703,PRA.87.052708}. In discussing the effect of interaction on $\xi_{\mathbf{Q}}$, we will be interested in (i) $\frac{\partial \xi_{\mathbf{Q}}}{\partial a^{-1}_{s}}$ and (ii) $\Delta_{\mathbf{Q}\mathbf{Q}^{\prime}}\equiv\xi_{\mathbf{Q}^{\prime}}-\xi_{\mathbf{Q}}$: The sign of the former reflects how $\xi_{\mathbf{Q}}$ for fixed $\mathbf{Q}$ changes with interaction, while that of the latter tells whether a large or small $\mathbf{Q}$ is energetically favored for a given interaction. Using Eq.~(\ref{Eq5}), we find $\Delta_{\mathbf{Q}\mathbf{Q}^{\prime}}
\simeq-[\int^{\infty}_{0}\frac{\gamma^{\varepsilon}_{\mathbf{Q}}}
{(\varepsilon-\varepsilon_{b})^2}d\varepsilon]^{-1}
\int^{\infty}_{0}\frac{\gamma^{\varepsilon}_{\mathbf{Q}^{\prime}}-
\gamma^{\varepsilon}_{\mathbf{Q}}}{\varepsilon-\varepsilon_{b}}
d\varepsilon$. Both of $\xi_{\mathbf{Q}}$ and $\Delta_{\mathbf{Q}\mathbf{Q}^{\prime}}$ rely crucially on $\gamma^{\varepsilon}_{\bf{Q}}$.
Thus, while the form of $\gamma^{\varepsilon}_{\bf{Q}}$ varies with specific setups [see Eq. (\ref{Eq4})], its qualitative analysis affords insights into generic behavior of $\xi_{\mathbf{Q}}$, as we elaborate next.
In order to give some analyses, we apply the further approximation
\begin{align}\label{Eq5a}
\xi_{\mathbf{Q}}&\simeq-\Big[\int^{\infty}_{0}\frac{\gamma^{\varepsilon}_{0}}{(\varepsilon-\varepsilon_{b})^2}d\varepsilon\Big]^{-1}
\int^{\infty}_{0}\frac{\gamma^{\varepsilon}_{\mathbf{Q}}-\gamma^{\varepsilon}_{0}}{\varepsilon-\varepsilon_{b}}
d\varepsilon\nonumber\\
&\propto\sqrt{-\varepsilon_b}\int^{\infty}_{0}\frac{\gamma^{\varepsilon}_{\mathbf{Q}}-\gamma^{\varepsilon}_{0}}{\varepsilon-\varepsilon_{b}}
d\varepsilon.
\end{align}

Consider first the simplest case where $\gamma^{\varepsilon}_{\mathbf{Q}}-\gamma^{\varepsilon}_{0}>0$ for all energy levels $\varepsilon$ \cite{foot2}, i.e., SOC induces an increase in the number of available scattering states at all energies. From Eq.~(\ref{Eq5a}), we see $\xi_\mathbf{Q}<0$, hence binding with finite $\mathbf{Q}$ leads to an energy decrease as compared to the SOC-free case, irrespective of the interacting strength.
Such energy drop, following from $\frac{\partial \xi_{\mathbf{Q}}}{\partial a^{-1}_{s}}>0$, can be further enhanced by increasing $a_s^{-1}$. If, moreover, $\gamma^{\varepsilon}_{\mathbf{Q}}$ increases monotonically with $\mathbf{Q}$, we have $\Delta_{\mathbf{Q}\mathbf{Q}^{\prime}}<0$, i.e., $\xi_{\mathbf{Q}}$ decreases with increasing $\mathbf{Q}$ for fixed scattering length. The amplitude of this decrease can be controlled by tuning the scattering length, which enhances with increased $a_s^{-1}$.

In contrast, if the effect of SOC is such that $\gamma^{\varepsilon}_{\mathbf{Q}}-\gamma^{\varepsilon}_{0}$ alters sign depending on the energy $\varepsilon$ of the state, $\xi_{\mathbf{Q}}$ can exhibit a very rich behavior. To demonstrate it, consider $\gamma^{\varepsilon}_{\mathbf{Q}}-\gamma^{\varepsilon}_{0}$ has opposite sign in the low- and high-energy regimes, with a sign flip occurring at the energy $\varepsilon_{0}$. Applying the mean value theorem to Eq.~(\ref{Eq5a}), we find
\begin{align}
\int^{\infty}_{0}\frac{\gamma^{\varepsilon}_{\mathbf{Q}}-\gamma^{\varepsilon}_{0}}
{\varepsilon-\varepsilon_{b}}d\varepsilon=f_{l}/(\varepsilon_1-\varepsilon_{b})+f_{h}/(\varepsilon_2-\varepsilon_{b}),
\end{align}
with $\varepsilon_{1}\in (0,\varepsilon_{0})$, and $\varepsilon_{2}\in
(\varepsilon_{0},\infty)$. Here $f_{l}=\int^{\varepsilon_{0}}_{0}
(\gamma^{\varepsilon}_{\mathbf{Q}}-\gamma^{\varepsilon}_{0})d\varepsilon$ and
$f_{h}=\int^{\infty}_{\varepsilon_{0}}
(\gamma^{\varepsilon}_{\mathbf{Q}}-\gamma^{\varepsilon}_{0})d\varepsilon$ are the number of scattering states in the low- and high-energy regimes, respectively. Since $f_l$ and $f_h$ have opposite signs, the contribution from the high-energy states to $\xi_{\mathbf{Q}}$ is suppressed by the smaller pre-factor compared to the low-energy states. Yet, such suppression becomes less significant when $a_s^{-1}$ increases, following similar reasoning as before. We thus expect the sign of $\xi_{\mathbf{Q}}$ to be mainly determined by the low-energy states for large $a_s$, whereas the high-energy states can become decisive for small $a_s$. This has interesting physical implications: by tuning the scattering length and hence the sign of $\xi_{\mathbf Q}$ and $\Delta_{\mathbf{Q}\mathbf{Q}^{\prime}}$, we can control whether a bound pair favors nonzero $\mathbf{Q}$, and even the specific choice of $\mathbf{Q}$.

\section{Typical behaviors of two-body ground states}
We now show that, combining $E^{\mathbf{Q}}_{th}$, above insights into the cooperative effects of interaction and SOC on $E_{\mathbf{Q}}^{sc}$ allows predictions on generic features of  the dispersion $E_{\bf{Q}}$. This can be best illustrated in two following cases.

\begin{figure*}[htbp]
  \centering
  \includegraphics[width=1.6\columnwidth]{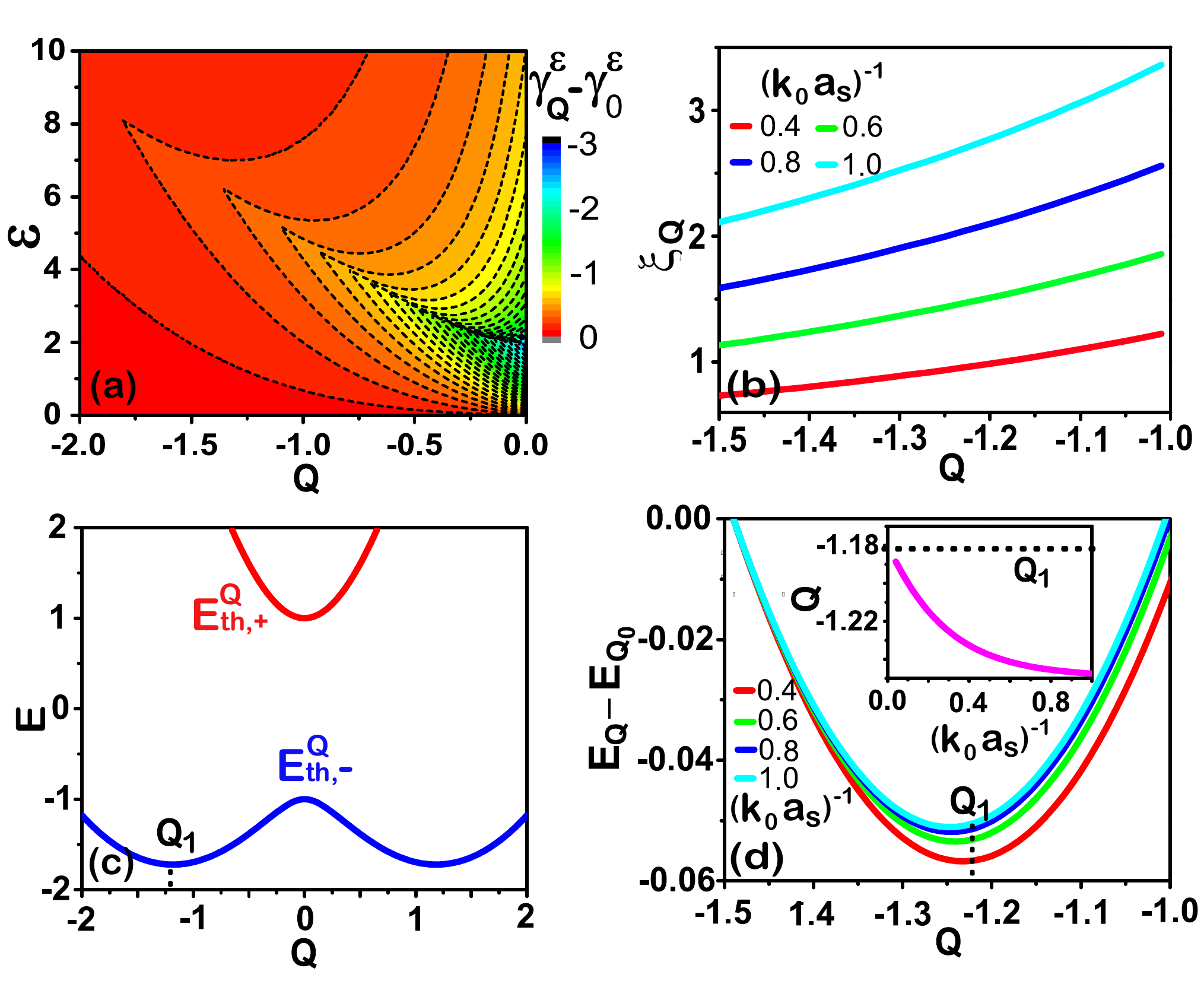}\\
  \caption{ Binding of spin-orbit coupled fermions in the vacuum. (a)
  The distribution of $\gamma^{\varepsilon}_{\mathbf{Q}}-\gamma^{\varepsilon}_{0}$.
  (b) $\xi_{\mathbf{Q}}$ as a function of $\mathbf{Q}$ with different
  $(k_0 a_s)^{-1}$ according to Eq.(\ref{Eq5}).
  (c) The helicity-dependent threshold energy $E^{\mathbf{Q}}_{th,+}$  ($E^{\mathbf{Q}}_{th,-}$) is the minimum energy of two particles with A in the upper (lower) helicity branch and
  a c.m. momentum $\mathbf{Q}$. (d) The two-body energy with different interacting strengths by exactly solving Eq. (\ref{Eq2}).
  The inset shows the variation of the ground state c.m. momentum. Here $\mathbf{Q}_{0}=-1.5k_0 e_x$. }\label{Fig1}
\end{figure*}

\begin{figure*}[htbp]
  \centering
  \includegraphics[width=1.6\columnwidth]{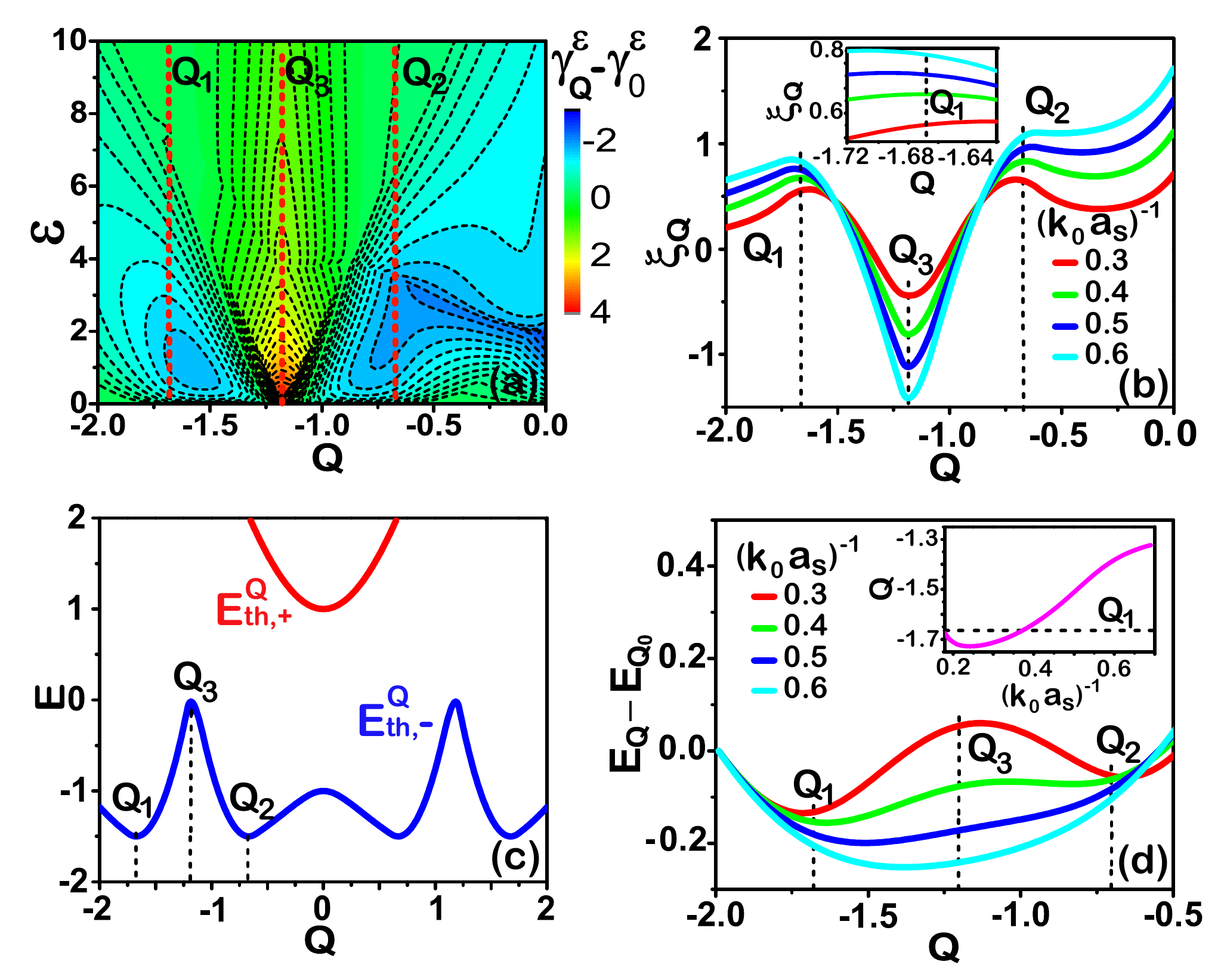}\\
  \caption{Binding of spin-orbit coupled fermions on top of a Fermi sea in single impurity system. (a)
  The distribution of $\gamma^{\varepsilon}_{\mathbf{Q}}-\gamma^{\varepsilon}_{0}$.
  (b) $\xi_{\mathbf{Q}}$ as a function of $\mathbf{Q}$ with different
  $(k_0 a_s)^{-1}$ according to Eq.(\ref{Eq5}). The inset shows $\xi_{\mathbf{Q}}$ in the region near $\mathbf{Q}_1$. (c) The threshold
 $E^{\mathbf{Q}}_{th}=\min{\{E^{\mathbf{Q}}_{th,-},E^{\mathbf{Q}}_{th,+}\}}$. The helicity-dependent threshold energy $E^{\mathbf{Q}}_{th,+}$  ($E^{\mathbf{Q}}_{th,-}$) is the minimum energy of two particles with A in the upper (lower) helicity branch and
a c.m. momentum $\mathbf{Q}$.
  (d) The two-body energy with different interacting strengths by exactly solving Eq. (\ref{Eq2}).
  The inset shows the variation of the ground state c.m. momentum. Here $\mathbf{Q}_{0}=-2k_0 e_x$. }\label{Fig2}
\end{figure*}

(i) If $E^{\mathbf{Q}}_{th}$ has only one minimum, without interaction, the two-body (molecular) ground state c.m. momentum $\mathbf{Q}_g$ will locate at $\mathbf{Q}_1$ where $E^{\mathbf{Q}}_{th}$ is minimized. By contrast, adding interaction can strongly modify $E_{\mathbf{Q}}^{sc}$ and thus $E_{\mathbf{Q}}$, according to previous analysis, which renders $\mathbf{Q}_g$ to deviate from $\mathbf{Q}_1$. Such deviation intimately depends on the behavior of $E_{\mathbf{Q}}^{sc}$: If $E_{\mathbf{Q}}^{sc}$ varies monotonically with
$\mathbf{Q}$ for a fixed scattering length, $\mathbf{Q}_g$ shifts from $\mathbf{Q}_1$ in such a way that a smaller $E_{\mathbf{Q}}^{sc}$ can be reached. Such shift can be further enhanced by increasing $a^{-1}_s$, provided it does not qualitatively alter the behavior of $E_{\mathbf{Q}}^{sc}$, i.e,. $E_{\mathbf{Q}}^{sc}$ stays increasing (or decreasing) with ${\bf Q}$ when varying $a^{-1}_s$ [c.f. inset of Fig.~\ref{Fig1}(d)]. If, instead, the behavior of $E_{\mathbf{Q}}^{sc}$ undergoes a qualitative change when $a_s^{-1}$ increases, e.g. from increasing to decreasing with ${\bf Q}$ [see inset of Fig.~\ref{Fig2}(c)], $\mathbf{Q}_g$ will first exhibit a zigzag away from $\mathbf{Q}_1$ before increasing above $\mathbf{Q}_1$ monotonically [see inset of Fig.~\ref{Fig2}(d)].

(ii) In general $E^{\mathbf{Q}}_{th}$ can have multiple local minima, each corresponding to a meta-stable state. For individual meta-stable state, the associated c.m. momentum exhibits similar behavior as in (i).  An interesting question then concerns how two-body (molecular) ground state transits among multiple meta-stable states when the interaction is tuned.
To address it, suppose for simplicity that $E^{\mathbf{Q}}_{th}$ has two degenerate local minima at $\mathbf{Q}_1$ and $\mathbf{Q}_2$ respectively, and $E_{\mathbf{Q}}^{sc}$ varies monotonically with
$\mathbf{Q}$ for a fixed scattering length. The two-body (molecular) ground state c.m. momentum $\mathbf{Q}$ is expected to be close to ${\bf Q}_1$ or ${\bf Q}_2$, depending on which corresponds to a smaller $E_{\mathbf{Q}}^{sc}$. If the behavior of $E_{\mathbf{Q}}^{sc}$ can be
changed qualitatively by tuning $a_s^{-1}$, say from increase to decrease
with $\mathbf{Q}$, a transition of the system between the two meta-stable states can be induced.
This phenomenon also occurs when the two local minima $E^{\mathbf{Q}}_{th}$
become non-degenerate, due to the competition between $E^{\mathbf{Q}}_{th}$ and $E_{\mathbf{Q}}^{sc}$, which is the origin of the transition discussed in Ref.~\cite{PRL.112.195301}. In addition, with the increasing of $a^{-1}_s$, $E_{\mathbf{Q}}^{sc}$ will dominate over $E_{th}^{\bf Q}$ in determining the dispersion of $E_{\bf Q}$. This may qualitative change the dispersion of two-body (molecular) energy, say
from a double-well type with two meta-stable states to a single-well type with one meta-stable state, which may cause the disappear of
the transition.

\section{Spin-orbit coupled three-component Fermi mixture}
Previous discussions from Sec. II to Sec. IV are not dependent on the concrete type of the SOC.
To give an example, below we present concrete calculations by solving Eq.~(\ref{Eq2}) for a system of interacting
Fermi mixture of $^{40}$K-$^{40}$K-$^{6}$Li
(A-A-B), where the atom $^{40}$K couples to SOC and the atom $^{6}$Li is spinless. Here, we choose an ($\alpha k_x\sigma_z+h\sigma_x$)-type SOC which can be readily realized experimentally in $^{40}$K \cite{ESOCF}.
In this three-component mixture, the $^{6}$Li
fermions are tuned close to a wide Feshbach resonance with
spin up species of $^{40}$K \cite{FM}.
The Hamiltonian for the system reads
\begin{align} \label{Eq6}
&H=\sum_{\mathbf{k},\sigma}\varepsilon^{a}_{\mathbf{k}}a^{\dagger}_{\mathbf{k},\sigma}a_{\mathbf{k},\sigma}
+\sum_{\mathbf{k}}(ha^{\dagger}_{\mathbf{k},\uparrow}a_{\mathbf{k},\downarrow}+
ha^{\dagger}_{\mathbf{k},\downarrow}a_{\mathbf{k},\uparrow})\nonumber\\
&+\sum_{\mathbf{k}}\varepsilon^{b}_{\mathbf{k}}b^{\dagger}_{\mathbf{k}}b_{\mathbf{k}}
+\frac{U}{V}\sum_{\mathbf{k,k^{\prime},q}}a^{\dagger}_{\mathbf{\frac{q}{2}+k},\uparrow}b^{\dagger}_{\mathbf{\frac{q}{2}-k}}
b_{\mathbf{\frac{q}{2}-k^{\prime}}}a_{\mathbf{\frac{q}{2}+k^{\prime}},\uparrow}\nonumber\\
&+\sum_{\mathbf{k}}(\alpha k_{x}a^{\dagger}_{\mathbf{k},\uparrow}a_{\mathbf{k},\uparrow}-\alpha k_{x}a^{\dagger}_{\mathbf{k},\downarrow}a_{\mathbf{k},\downarrow}).
\end{align}
Here $a_{\mathbf{k},\sigma}$ $(\sigma=\uparrow,\downarrow)$ denotes the annihilation operator of a SOC-free particle $A$ with spin $\sigma$ and momenta ${\bf k}$, while the operator $b_{\mathbf{k}}$ annihilates a particle $B$ with momenta ${\bf k}$. In addition, $\varepsilon^{a(b)}_{\mathbf{k}}={k^2}/{(2m_{a(b)})}$ is the kinetic energy of particle $A(B)$. The SOC parameters $h$ and $\alpha$ are respectively proportional to the Raman coupling strength and the momentum transfer in the Raman process generating the SOC \cite{ESOCF}. We also note that via a global pseudo-spin rotation such SOC can be transformed to an equal weight combination of Rashba-type and Dresselhaus-type SOC ($\alpha k_x\sigma_y+h\sigma_z$) which is the first SOC generated in ultracold atomic gases \cite{ESOCB}. Therefore, the ($\alpha k_x\sigma_z+h\sigma_x$)-type SOC can be
interpreted as an equal weight combination of Rashba-type and Dresselhaus-type SOC \cite{ESOCF}.

In the presence of SOC, the single-particle eigenstates of $A$ in the helicity basis are created by operators $a^{\dagger}_{\mathbf{k},\pm}
=\lambda^{\pm,\uparrow}_{\mathbf{k}}a^{\dagger}_{\mathbf{k},\uparrow}+\lambda^{\pm,\downarrow}_{\mathbf{k}}%
a^{\dagger}_{\mathbf{k},\downarrow}$, with $\lambda^{\pm,\uparrow}_{\mathbf{k}}=\pm\zeta^{\pm}_{\mathbf{k}}$,
$\lambda^{\pm,\downarrow}_{\mathbf{k}}=\zeta^{\mp}_{\mathbf{k}}$, and $\zeta^{\pm}_{\mathbf{k}}=
[\sqrt{h^2+\alpha^2 k^2_{x}}\pm\alpha k_x]^{1/2}/\sqrt{2}[h^2+\alpha^2 k^2_{x}]^{1/4}$, with $+(-)$ labelling the upper (lower) helicity branch.
The single particle dispersions of two helicity branches are
$\varepsilon^{a}_{\mathbf{k},\pm}=\varepsilon^{a}_{\mathbf{k}}\pm\sqrt{h^2+\alpha^2 k^2_{x}}$. Here we have measured the energy in the unit of $E_{0}=2\alpha^2 m_a/\hbar^2$, the momentum in the unit of $k_{0}=2\alpha m_a/\hbar^2$, and $h=0.4E_{0}$.

We first present our results for the binding of $A$ and $B$ in the vacuum, as summarized in Fig.~\ref{Fig1}. The density of states [see Fig.~\ref{Fig1}(a)] exhibits a monotonic decrease with both ${\bf Q}$ and $\varepsilon$. As expected, $E^{sc}_{\mathbf Q}$ will change monotonically with respect to both ${\bf Q}$ and $a_s^{-1}$ [see
Fig.~\ref{Fig1}(b)]. Together with $E^{\mathbf Q}_{th}$ [see Fig.~\ref{Fig1}(c)], we see that the actual ground
state c.m. momenta will be pulled to the direction with a smaller magnitude than $\mathbf{Q}_1$ and the increase of $a^{-1}_s$
will enhance this tendency [see Fig.~\ref{Fig1}(d)].

\begin{figure*}[htbp]
  \centering
  \includegraphics[width=1.6\columnwidth]{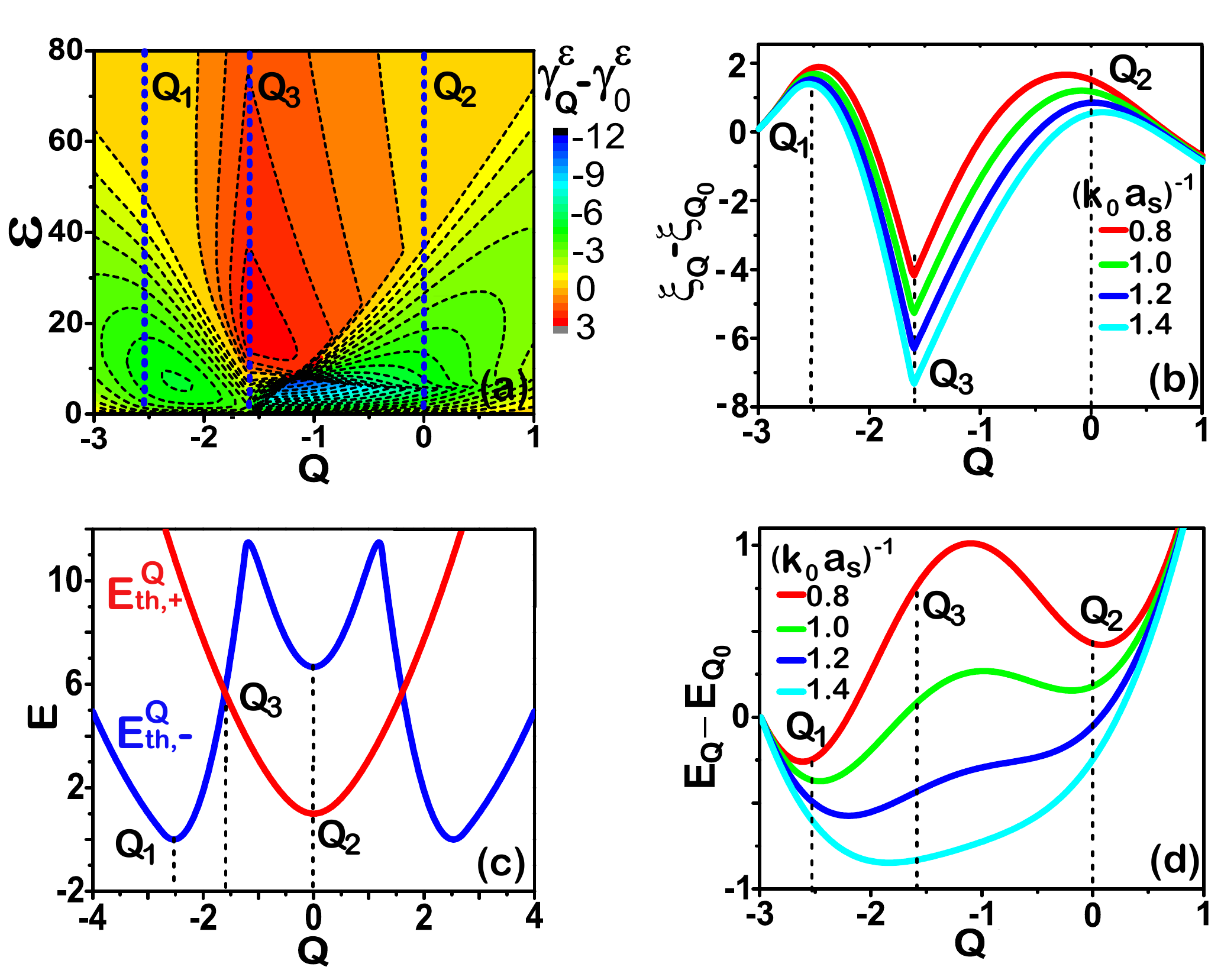}\\
  \caption{(a)
  The distribution of $\gamma^{\varepsilon}_{\mathbf{Q}}-\gamma^{\varepsilon}_{0}$.
  (b) $\xi_{\mathbf{Q}}$ as a function of $\mathbf{Q}$ with different
  $(k_0 a_s)^{-1}$ according to Eq.(\ref{Eq5}).  (c) The threshold energy $E^{\mathbf{Q}}_{th}=\min{\{E^{\mathbf{Q}}_{th,-},E^{\mathbf{Q}}_{th,+}\}}$. The helicity-dependent threshold energy $E^{\mathbf{Q}}_{th,+}$  ($E^{\mathbf{Q}}_{th,-}$) is the minimum energy of two particles with A in the upper (lower) helicity branch and
a c.m. momentum $\mathbf{Q}$. (d) The two-body energy with different interacting strengths by exactly solving Eq. (\ref{Eq2}).
   Here $\mathbf{Q}_{0}=-3k_0 e_x$.}\label{Fig3}
\end{figure*}

We now turn to the binding of $A$ and $B$ on the top of the Fermi sea of $A$ in the situation where a single impurity of $B$ immerses in a non-interacting Fermi sea of spin-orbit coupled $A$ with the Fermi energy $E_{h}=-1.5E_0$, as illustrated in Fig.~\ref{Fig2}. There, both the density of states [see Fig.~\ref{Fig2}(a)] and $E^{sc}_{\mathbf{Q}}$ [see Fig.~\ref{Fig2}(b)] exhibit a rich behavior. In addition, from $E^{\mathbf Q}_{th}$ in Fig.~\ref{Fig2}(c), we see that there exist two meta-stable
states near $\mathbf{Q}_1$ and $\mathbf{Q}_2$, respectively. Let us first analyze the c.m. momenta associated with the meta-stable states, e.g.,
the one formed near $\mathbf{Q}_1$. Seen from Fig.~\ref{Fig2}(a), $\gamma^{\varepsilon}_{\mathbf{Q}}$ for c.m. momentum near $\mathbf{Q}_1$ decreases with
$\mathbf{Q}$ in the low energy region (e.g. $0<\varepsilon<2E_0$), but increases in the
high energy region (e.g. $6 E_0<\varepsilon<10 E_0$). In addition, near $\mathbf{Q}_1$, $E^{\mathbf{Q}}_{sc}$ [see the inset of Fig.~\ref{Fig2}(b)] shows a qualitative change with increasing of $a^{-1}_s$. We thus expect from earlier discussions a zigzag behavior of c.m. momenta of the meta-stable state, as confirmed by our results plotted in the inset of Fig.~\ref{Fig2}(d). Next, we discuss which of the two meta-stable states is energetically favored. Due to the degeneracy of the two local minima
of $E^{\mathbf{Q}}_{th}$, this is determined by the density of states, which is larger near $\mathbf{Q}_1$ than that near
$\mathbf{Q}_2$ [see Fig.~\ref{Fig2}(a)]. Hence the meta-stable state near $\mathbf{Q}_1$ is energetically favored by $E^{\mathbf{Q}}_{sc}$ [see
Fig.~\ref{Fig2}(b)]. We thus expect the molecular ground state c.m. momentum to be near $\mathbf{Q}_1$, well agreeing with Fig.~\ref{Fig2}(d).

Comparing the binding of $A$ and $B$ in the vacuum and on top of the filled Fermi sea, we observe that the presence of Fermi sea not only elevates $E^{\mathbf{Q}}_{th}$ in the regime $\mathbf{Q}_1 < \mathbf{Q}
< \mathbf{Q}_2$, giving rise to two minima, but also enhances the density of states there. Consequently, the minimum of
$E^{\mathbf{Q}}_{sc}$ occurs at $\mathbf{Q}_3$, and the two meta-stable states merge together [see Fig.~\ref{Fig2}(d)] following from previous analysis. We remark that, while the SOC here has dramatically modified the density of states compared to the SOC-free case, our analyses based on perturbation treatment agree remarkably well with the exact numerical results.

\section{Concluding discussions and Summary}
When the Fermi sea has only one
Fermi surface, the two meta-stable states formed near $\mathbf{Q}_1$ and
$\mathbf{Q}_2$ are favored by the threshold energy and the density of
states, respectively, see Fig. \ref{Fig3}. In Ref. \cite{PRL.112.195301} with
high Fermi energy, tuning the interaction can induce a transition
between the two meta-stable states. In contrast, as illustrated in
Fig. \ref{Fig3} where the Fermi energy $E_h=0$, such transition is
missing and the increase of $a^{-1}_s$ will eventually cause a merge
of the two meta-stable states. With an increase of the Fermi energy,
our case crossovers to that discussed in
Ref. \cite{PRL.112.195301}. In addition, we note that for
the single impurity Fermi system we only consider the lowest energy state within our ansatz, the ground state of the system should be given by connecting the molecular ground state to the
polaron ground state which describes the particle-hole
excitations above the Fermi sea.

Summarizing, we have investigated how the tuning of interacting
strength of an attractive s-wave interaction affects two-body energy under
certain distribution of the density of states.
Combining with the dispersion of the threshold energy,
we can predict typical behavior of the two-body bound state when tuning the scattering length and hence the interaction, including the change of the c.m. momentum of the two-body ground state
and the competition between multiple meta-stable states. Our perturbation analyses are not dependent on the
concrete type of SOC and corroborated by the exact numerical solution of the two-body problem for a spin-orbit coupled Fermi
mixture of $^{40}$K-$^{40}$K-$^6$Li, even though the density of states is significantly altered by the effect of SOC.

%
\section{Acknowledgments}
We thanks Ying Hu for helpful discussion. The work is supported by the National Basic Research
Program of China (973 Program) (Grants No. 2013CBA01502
and No. 2013CB834100), the National Natural Science Foundation
of China (Grants  No. 11374040, No.
11475027, No. 11575027, and No. 11274051).


\begin{thebibliography}{99}

%
%
%
%
%
%
%
%
%
%
%
%
%
%
%
%
%
%
%



\bibitem{TSOC1}  Jean Dalibard, Fabrice Gerbier, Gediminas Juzeli\={u}nas, and Patrik \"{O}hberg, Rev. Mod. Phys. \textbf{83}, 1523 (2011).
\bibitem{ESOCB}  Y.-J. Lin, K. Jimen\'{e}z-Garc\'{i}a, and I. B. Spielman, Nature (London) \textbf{471}, 83 (2011).
\bibitem{ESOCB1} Jin-Yi Zhang, Si-Cong Ji, Zhu Chen, Long Zhang, Zhi-Dong Du, Bo Yan, Ge-Sheng Pan, Bo Zhao, YouJin Deng, Hui Zhai, Shuai Chen, and Jian-Wei Pan, Phys. Rev. Lett. \textbf{109}, 115301 (2012).
\bibitem{ESOCF} Pengjun Wang, Zeng-Qiang Yu, Zhengkun Fu, Jiao Miao, Lianghui Huang, Shijie Chai, Hui Zhai, and Jing Zhang, Phys. Rev. Lett. \textbf{109}, 095301 (2012).
\bibitem{ESOCF1} Lawrence W. Cheuk, Ariel T. Sommer, Zoran Hadzibabic, Tarik Yefsah, Waseem S. Bakr, and Martin W. Zwierlein, Phys. Rev. Lett. \textbf{109}, 095302 (2012); Lianghui Huang, Zengming Meng, Pengjun Wang, Peng Peng, Shao-Liang Zhang, Liangchao Chen, Donghao Li, Qi Zhou and Jing Zhang, Nat. Phys. \textbf{10}, 1038 (2016).
\bibitem{ZH1} H. Zhai, Int. J. Mod. Phys. B \textbf{26}, 1230001 (2012).
\bibitem{ISB} V. Galitski and I. B. Spielman, Nature, \textbf{494}, 49 (2013).
\bibitem{ZH2} H. Zhai, Rep. Prog. Phys., \textbf{78}, 026001 (2015).


%



\bibitem{PRB.83.094515} Jayantha P. Vyasanakere, and Vijay B. Shenoy, Phys. Rev. B \textbf{83}, 094515 (2011).
\bibitem{PRB.84.014512} Jayantha P. Vyasanakere, Shizhong Zhang, and Vijay B. Shenoy, Phys. Rev. B
\textbf{84}, 014512 (2011). 
\bibitem{PRL.107.195304} Hui Hu, Lei Jiang, Xia-Ji Liu, and Han Pu, Phys. Rev. Lett. \textbf{107},
195304 (2011). 
\bibitem{PRL.107.195305} Zeng-Qiang Yu and Hui Zhai, Phys. Rev. Lett. \textbf{107}, 195305 (2011). 
\bibitem{PRA.87.033629}  Ren Zhang, Fan Wu, Jun-Rong Tang, Guang-Can Guo, Wei Yi, and Wei Zhang, Phys. Rev. A. \textbf{87}. 033629 (2013). 
\bibitem{PRA.88.033609}  Vijay B. Shenoy, Phys. Rev. A. \textbf{88}. 033609 (2013). 
\bibitem{PRA.89.063610}  Fan Wu, Ren Zhang, Tian-Shu Deng, Wei Zhang, Wei Yi, and Guang-Can Guo, Phys. Rev. A. \textbf{89}.063610 (2014). 
\bibitem{PRL.112.195301} Lihong Zhou, Xiaoling Cui, and Wei Yi, Phys. Rev. Lett. \textbf{112}, 195301 (2014). 

\bibitem{c9} F. Chevy, Phys. Rev. A \textbf{74}, 063628 (2006).
\bibitem{c10} R. Combescot, A. Recati, C. Lobo and F. Chevy, Phys. Rev. Lett, \textbf{98}, 180402 (2007).
\bibitem{c11} Sascha Zollner, G. M. Bruun, and C. J. Pethick, Phys. Rev. A \textbf{83}, 021603(R) (2011).
\bibitem{c12} Marco Koschorreck, Daniel Pertot, Enrico Vogt, Bernd Frohlich, Michael Feld and Michael Kohl, Nature \textbf{485}, 619 (2012).
\bibitem{c13} Wei Yi and Wei Zhang, Phys. Rev. Lett \textbf{109}, 140402 (2012).


\bibitem{RMP.82.1225} Cheng Chin, Rudolf Grimm, Paul Julienne, and Eite Tiesinga, Rev. Mod. Phys. \textbf{82}, 1225 (2010).
\bibitem{FM} Andr\'{e} Schirotzek, Cheng-Hsun Wu, Ariel Sommer, and Martin W. Zwierlein, Phys. Rev. Lett. \textbf{102}, 230402 (2009); M. Koschorreck, D. Pertot, E. Vogt, B. Fr\"{o}hlich, M. Feld, and M. K\"{o}hl, Nature (London) \textbf{485}, 619 (2012).

\bibitem{RdNC.31.247} W. Ketterle and M. W. Zwierlein,
Rivista del Nuovo Cimento \textbf{31}, 247-422 (2008).
\bibitem{foot}{We note that for homo-nuclear systems, our derivation for $\gamma^{\varepsilon}_{\mathbf{Q}}$ reduces to the
singlet density of states as discussed in Ref. \cite{PRA.88.033609}.  }
\bibitem{Sup1} With Eq.(\ref{Eq3}) and renomolization eqaution, we have $\int^{\infty}_{0}
\frac{\gamma^{\varepsilon}_{0}}{\varepsilon_{b}-\varepsilon}d\varepsilon=\int^{\infty}_{0}
\frac{\gamma^{\varepsilon}_{\mathbf{Q}}}{\xi_{\mathbf{Q}}+\varepsilon_{b}-\varepsilon}d\varepsilon$. The right hand of the equation
can be writen as follow $\int^{\infty}_{0}
\frac{\gamma^{\varepsilon}_{\mathbf{Q}}}{\xi_{\mathbf{Q}}+\varepsilon_{b}-\varepsilon}d\varepsilon=
\int^{\infty}_{0}
[\frac{\gamma^{\varepsilon}_{\mathbf{Q}}}{\varepsilon_{b}-\varepsilon}-
\frac{\xi_{\mathbf{Q}}\gamma^{\varepsilon}_{\mathbf{Q}}}{(\varepsilon_{b}-\varepsilon)^2}+
\frac{\xi_{\mathbf{Q}}^2\gamma^{\varepsilon}_{\mathbf{Q}}}{(\varepsilon_{b}-\varepsilon)^3}+...]d\varepsilon$.
To first order of $\xi_{\mathbf{Q}}$ ($|\frac{\varepsilon_{\mathbf{Q}}}{\varepsilon_{b}}|<<1$), we arrive Eq.(\ref{Eq5}).
\bibitem{PRA.85.022705} Xiaoling Cui, Phys. Rev. A \textbf{85}, 022705 (2012).
\bibitem{PRA.86.042707} Peng Zhang, Long Zhang, and Wei Zhang, Phys. Rev. A \textbf{86}, 042707 (2012).

\bibitem{PRA.86.053608} Peng Zhang, Long Zhang, and Youjin Deng, Phys. Rev. A \textbf{86}, 053608 (2012).

\bibitem{PRA.87.032703} Yuxiao Wu and Zhenhua Yu, Phys. Rev. A, \textbf{87}, 032703 (2013).
\bibitem{PRA.87.052708} Hao Duan, Li You, and Bo Gao, Phys. Rev. A, \textbf{87}, 052708 (2013).






\bibitem{foot2}{the analysis for the case with $\gamma^{\varepsilon}_{\mathbf{Q}}-\gamma^{\varepsilon}_{0}>0$ is similar.}
\end{thebibliography}
\end{document}